\documentclass[a4paper,11pt]{article}
\usepackage{pos}

\usepackage[T1]{fontenc}
\usepackage[utf8]{inputenc}
\usepackage[american]{babel}
\usepackage{mathtools}
\usepackage{upgreek}
\usepackage{dsfont}

\newcommand{\cpa}{\ensuremath{k_{\parallel}^{0}}}
\newcommand{\apar}{\ensuremath{a_{\parallel}}}
\newcommand{\bpar}{\ensuremath{b_{\parallel}}}
\newcommand{\aperp}{\ensuremath{a_{\perp}}}
\newcommand{\bperp}{\ensuremath{b_{\perp}}}

\title{Constraining the global heliospheric transport of galactic cosmic rays in solar cycles 23 and 24}
\ShortTitle{Global transport of GCRs in SCs 23 and 24}

\author*[a,b]{Claudio Corti}
\author[c]{Peter Sadowski}
\author[a]{Nikolay Nikonov}
\author[d]{Marius Potgieter}
\author[a]{Veronica Bindi}

\affiliation[a]{University of Hawaii at Manoa, Physics and Astronomy Department,\\
  2505 Correa Road, Honolulu HI, USA}

\affiliation[b]{NASA Goddard Space Flight Center, CCMC,\\
   8800 Greenbelt Road, Greenbelt MD, USA}

\affiliation[c]{University of Hawaii at Manoa, Information and Computer Science Department,\\
   1680 East-West Road, Honolulu HI, USA}

\affiliation[d]{Christian-Albrecths University, Institute for Experimental and Applied Physics,\\
   Leibniztra{\ss}e 11-19, Kiel, Germany}

\emailAdd{corti@hawaii.edu}
\emailAdd{peter.sadowski@hawaii.edu}
\emailAdd{nikonov@hawaii.edu}
\emailAdd{marius.s.potgieter@gmail.com}
\emailAdd{bindi@hawaii.edu}

\abstract{
   Galactic cosmic rays (GCRs) are affected by solar modulation while they propagate through the heliosphere.
   The study of the time variation of GCR spectra observed at Earth can shed light on the underlying physical processes, specifically diffusion and particle drifts.
   We combine a state-of-the art 3D numerical model of GCR transport in the heliosphere with a neural-network-accelerated Markov chain Monte Carlo to constrain the rigidity and time dependence of the global transport coefficients, using precise GCR data from the PAMELA and AMS-02 experiments between 2006 and 2019.
}

\ConferenceLogo{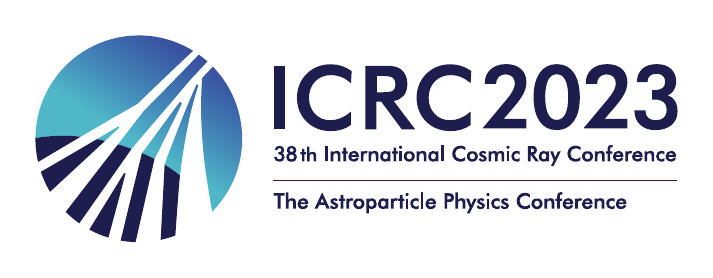}

\FullConference{%
38th International Cosmic Ray Conference (ICRC2023)\\
  26 July - 3 August, 2023\\
  Nagoya, Japan}

\begin{document}
\maketitle

\section{Introduction}
Galactic cosmic rays (GCRs) constitute a major radiation hazard for deep-space human exploration to the Moon, Mars, and beyond.
GCRs vary on different time scales: from a few days, driven by short-term solar wind disturbances, to the well known 11- and 22-year solar cycles, to even longer periodicities.
Studying the time dependence of GCRs is paramount to a better understanding of the origin of the solar activity and to predicting variations in the space environment.
The amount of data collected and of theoretical developments since the beginning of the space age allowed for great progress in understanding the spatial and time variation of GCRs, especially in the vicinity of Earth.
However, a number of key open questions such as the nature of the global heliospheric turbulence still remain, inhibiting our ability to make physics-based long-term predictions of radiation levels outside a spacecraft.

After entering the heliosphere, GCRs are advected by the solar wind, scatter on the heliospheric magnetic field (HMF) irregularities, drift along the HMF curvature, gradients, and the wavy heliospheric current sheet (HCS), adiabatically lose/gain energy due to solar wind expansions/contractions, and experience shocks at various boundaries~\citep{bib:parker65:modulation,bib:potgieter13:solar-modulation}.
The relation between the HMF turbulence and diffusion coefficients, in particular the rigidity behavior of the parallel and perpendicular mean free paths, depends on the chosen diffusion theory and turbulence geometry~\citep{bib:jokipii66:qlt,bib:matthaeus03:nlgct}.

Here we analyze the long-term variations of GCR protons near Earth measured directly in space during solar cycles 24 and 25 by the PAMELA and AMS-02 experiments.
Monthly averaged data are fit to spectra obtained from a 3D steady-state finite-difference model~\citep{bib:potgieter14:modulation-galactic-protons,bib:vos15:pamela-modulation,bib:corti19:numericalmodeling,bib:bisschoff19:new-very-local}, using a Markov Chain Monte Carlo (MCMC) technique to estimate a posterior probability density function (PDF) over the free parameters of the model.
We speed up the MCMC inference by training a neural network (NN) that approximates the numerical model output, using a coarse grid of known solutions as training data to predict the GCR spectra from model parameters.

\section{Numerical model for GCR propagation in the heliosphere}

\subsection{Heliospheric magnetic field, current sheet, and diffusion tensor}
The HMF implemented in this model is the Parker field as modified by~\cite{bib:smith91:hmf-sb-mod}, while the HCS is implemented as in~\cite{bib:kota83:drift-3d-model}.

The rigidity dependence of the parallel diffusion coefficient (DC) is approximated by a double power-law with a smooth change of slope, while the radial dependence is assumed to be inversely proportional to the magnitude of the HMF:
\begin{equation} \label{eqn:numerical-model:kpar}
   k_{\parallel} = \cpa \beta \frac{1\,\mathrm{nT}}{B} \left( \frac{R}{R_{k}} \right)^{a} \left[ 1 + \left( \frac{R}{R_{k}} \right)^{s} \right]^{\frac{b-a}{s}},
\end{equation}
where $\cpa$ is a normalization factor, $R_{k}$ is the rigidity at which the transition between the two power-laws happens, $a$ and $b$ are, respectively, the slopes of the low- and high-rigidity power-laws, and $s$ controls the smoothness of the transition.
This parametrization reproduces the rigidity dependence predicted by quasi-linear theory\citep{bib:jokipii66:qlt}.
The perpendicular diffusion coefficients are assumed to be proportional to the parallel diffusion coefficient, $k_{\perp,r} = k_{\perp,r}^{0} k_{\parallel}$ and $k_{\perp,\theta} = u(\theta) k_{\perp,\theta}^{0} k_{\parallel}$, where $k_{\perp,r}^{0}$ and $k_{\perp,\theta}^{0}$ are scaling factors of the order of percent, while $u(\theta)$ is a smooth transition function that enhances the perpendicular diffusion in the polar regions, tuned to reproduce cosmic ray observations at higher latitudes by the \textit{Ulysses} spacecraft~\citep{bib:heber06:modulation}.
The rigidity slopes of the parallel and perpendicular diffusion coefficients are not constrained to be the same, as expected by the non-linear guiding center theory~\citep{bib:matthaeus03:nlgct}.

The drift coefficient is defined as:
\begin{equation} \label{eqn:numerical-model:drift}
   k_{A} = k_{A}^{0} \frac{\beta R}{3 B} \frac{\left( R/R_{A} \right)^{2}}{1 + \left( R/R_{A} \right)^{2}},
\end{equation}
where $k_{A}^{0}$ is a normalization factor that can be used to reduce the overall drift effects, while $R_{A}$ is the rigidity below which the drift is suppressed due to scattering~\citep{bib:minnie07:drift-scattering}.

\subsection{Solar wind}
The solar wind velocity profile is assumed to be separable in a radial and latitudinal component: $\mathbf{V}_{sw}(r,\,\theta) = V_{0}V_{r}(r)V_{\theta}(\theta)\hat{\mathbf{r}}$.
The radial component describes the fast rise to supersonic speed within the first 0.3 AU from the Sun and the transition to subsonic speed at the termination shock, while the latitudinal term describes the transition between the slow (equatorial) and fast (polar) component of the solar wind:
\begin{spreadlines}{1ex}
   \begin{equation} \label{eqn:numerical-model:vsw-lat-asym}
      V_{\theta}(\theta) =
      \begin{dcases}
         \frac{V_{N} + V_{eq}}{2} - \frac{V_{N} - V_{eq}}{2} \tanh \left[ 3 \left( \theta' + \delta \right)\right],\ 0 < \theta < \pi/2\\
         \frac{V_{S} + V_{eq}}{2} + \frac{V_{S} - V_{eq}}{2} \tanh \left[ 3 \left( \theta' - \delta \right)\right],\ \pi/2 < \theta < \pi
      \end{dcases}
   \end{equation}
\end{spreadlines}
where $V_{N}$, $V_{S}$, and $V_{eq}$ are, respectively, the North-pole, South-pole, and equatorial solar wind speed components, $\theta' = \theta - \pi/2$, and $\delta$ is the angle at which the transition between the equatorial and polar streams begins, here set to be equal to the tilt angle in each analyzed time interval.

The parametrization of the latitudinal dependence of the solar wind in Equation \ref{eqn:numerical-model:vsw-lat-asym} is based on \textit{Ulysses} measurements during the three fast latitude scans in 1994--1995, 2000--2001, and 2007--2008~\citep{bib:mccomas03:threedimensionalsolar}.
However, these observations don't cover the full period of the PAMELA and AMS-02 data (2006 -- 2019).
Here we rely on the solar wind latitudinal structure inferred by IPS observations conducted by the Institute for Space-Earth Environmental Research in Nagoya, Japan~\citep{bib:tokumaru10:ips,bib:tokumaru12:ips} following the methodology proposed by~\cite{bib:sokol13:heliolatitudetime}\footnote{An updated dataset, extending the one in~\cite{bib:sokol13:heliolatitudetime} from 1985 to 2019, was obtained from Justyna M. Sokó\l\ (private communication, August 2019).}.
The availability of \textit{in-situ} data in the ecliptic (OMNI) and outside of the ecliptic (\textit{Ulysses}) allows to cross-calibrate IPS measurements, yielding an average uncertainty of about 50 km/s.

\begin{figure}[!h]
   \centering
   \includegraphics[width=\textwidth]{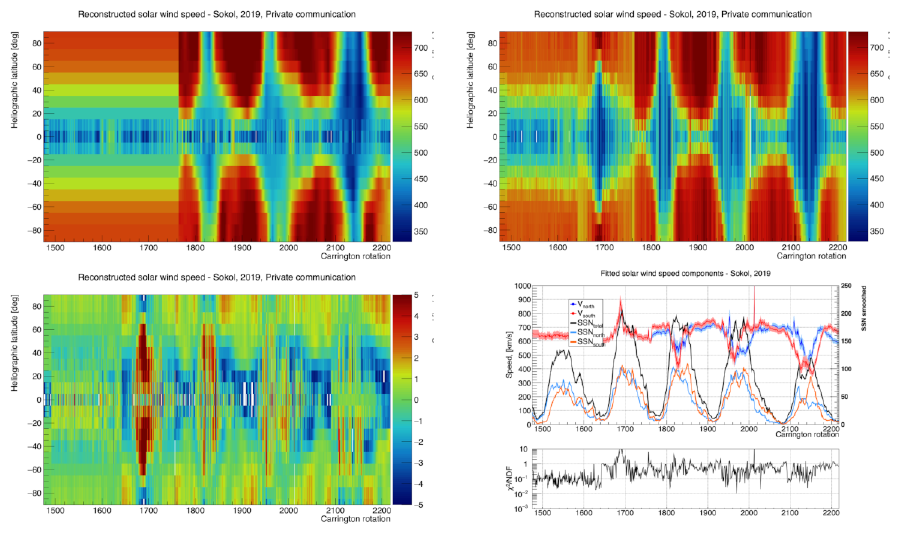}
   \caption{
      \textbf{Top left:} Solar wind latitudinal profile derived from IPS measurements.
      \textbf{Top right:} Reconstructed solar wind latitudinal profile (Eq. \ref{eqn:numerical-model:vsw-lat-asym}).
      \textbf{Bottom left:} Pull distribution of the reconstructed solar wind latitudinal profile.
      \textbf{Bottom right:} Fitted North and South solar wind speeds, together with the total and hemispheric CR-averaged sunspot number.
      The normalized $\chi^{2}$ is shown in the small panel below.
   }
   \label{fig:solar-wind-fits}
\end{figure}

Figure \ref{fig:solar-wind-fits} shows a comparison of our latitudinal solar wind profile (top right) with the one derived from IPS measurements (top left).

\section{Bayesian inference}
For each observed time interval (Carrington rotations for PAMELA, Bartels rotations for AMS-02), we fix the value of tilt angle, HMF at Earth, and polar solar wind speed to their 1-year backward average.
We use a MCMC sampling strategy to infer a PDF over the model free parameters: \cpa, \apar, \bpar, \aperp, \bperp.
If $\boldsymbol{\theta}$ is the vector of the model parameters, then $P(\boldsymbol{\theta}|\mathrm{data})$ is the posterior PDF of the parameters conditioned to the observed data, which can be calculated with Bayes theorem as:
\begin{equation}
P(\boldsymbol{\theta}|\mathrm{data}) = \frac{P(\mathrm{data}|\boldsymbol{\theta})\, P(\boldsymbol{\theta})}{P(\mathrm{data})},
\end{equation}
where $P(\mathrm{data}|\boldsymbol{\theta})$ is the probability of the observed data conditioned to a specific set of parameters (likelihood), $P(\boldsymbol{\theta})$ is the prior probability of the parameters, and $P(\mathrm{data})$ is a normalization factor that does not depend on the parameters.
The likelihood is defined as $\exp(-\chi^{2}(\boldsymbol{\theta})/2)$, where $\chi^{2}$ is the standard chi-squared between a model solution and a specific dataset.

In \cite{bib:zhang17:hamiltonian-monte-carlo} a NN surrogate model is used to speed up intermediate evaluations of Hamiltonian Monte Carlo (HMC).
Here, instead, we use a NN surrogate model to perform \textit{all} the likelihood evaluations.
This is motivated by two facts: (1) the neural network is faster to evaluate than the numerical simulation; and (2) the neural network can be used to compute gradients for HMC.
We use a NN trained using the \textsc{elegy}~\citep{bib:elegy2020repository} package for \textsc{jax}~\citep{bib:jax2018github}.
The model has three hidden layers of SELU (scaled exponential linear unit) activation~\citep{bib:klambauer17:self-normalizing-NN} and a linear output layer.
The input layer corresponds to model parameters used to describe the heliosphere (solar magnetic polarity, tilt angle, HMF intensity at 1 AU, and solar wind polar speed) and the DC (normalization of parallel DC, low- and high-rigidity slopes for parallel and perpendicular DCs).
The output layer corresponds to the rigidity spectrum at 1 AU in 32 steps from 0.2 to 200 GV.

HMC samples are generated using the No U-Turn Sampler (NUTS)~\citep{bib:hoffman14:nouturn-sampler-adaptively} implemented in Tensorflow Probability~\citep{bib:dillon17:tensorflow-distributions}.
We note that the diffusion coefficient parameters are constrained to be in a physical range.
This was done using an unnormalized prior distribution that is uniform in the domain of the training data, and exponentially decays in every direction outside that domain.
This was effective in preventing the HMC from wandering beyond the input region for which we trust the NN, but it is an artificial constraint on the HMC.
Autocorrelation plots show that the generated samples have very little correlation after 50 steps.
Thus, no thinning was used for our sample.

\begin{figure}[!h]
   \centering
   \includegraphics[width=0.96\textwidth]{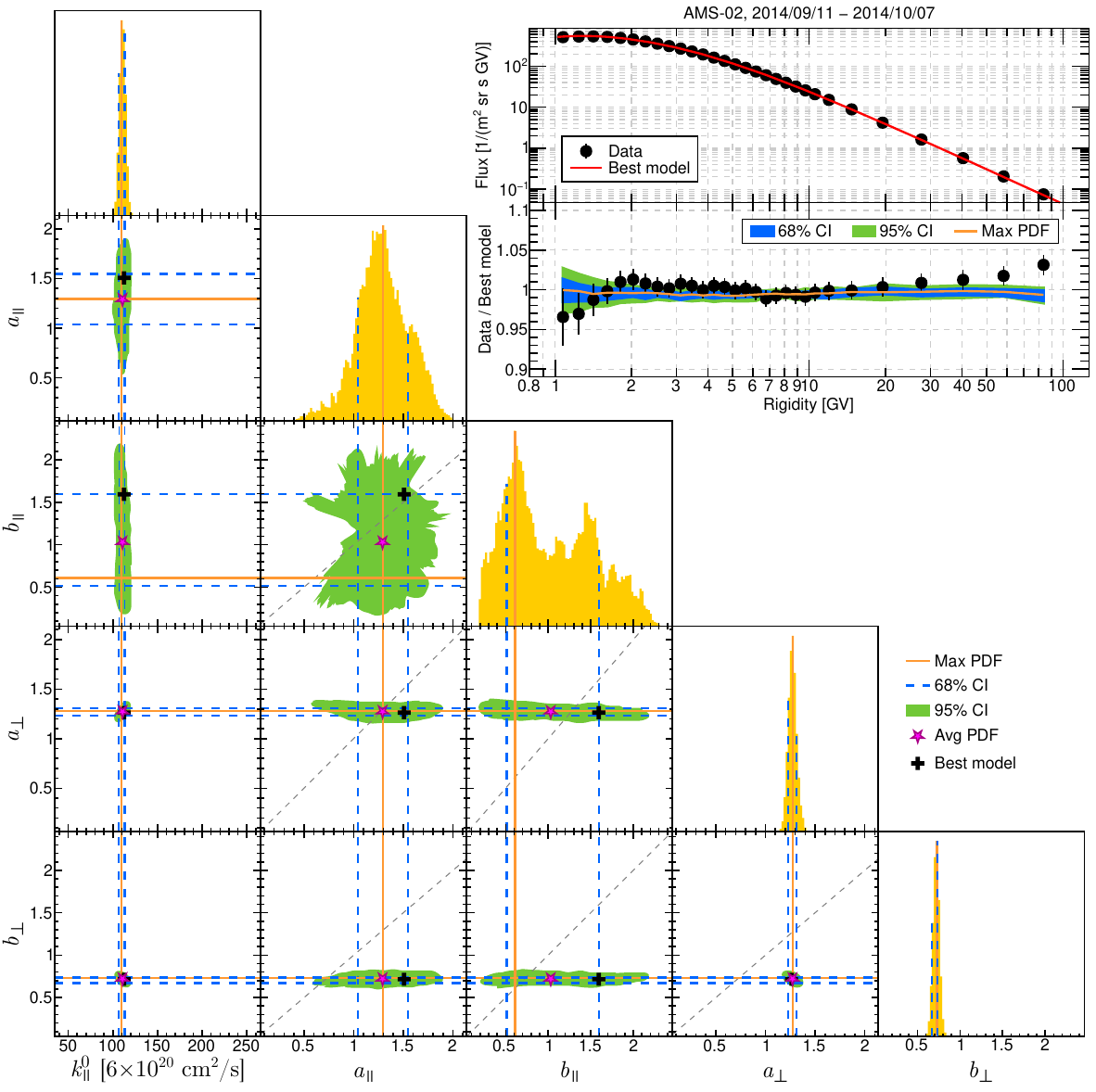}
   \caption{
      2D and 1D PDFs of the diffusion coefficient parameters obtained from the HMC for GCR protons measured by AMS-02 in 2014/09/11--2014/10/07.
      The top right panel shows a comparison of the maximum likelihood NN model with observations, together with the 68\% and 95\% credible intervals.
   }
   \label{fig:HMC-distributions}
\end{figure}

Figure \ref{fig:HMC-distributions} shows an example of the PDFs of the free parameters obtained with the HMC.

The PDF is very narrow for the normalization of the DC and for the slopes of the perpendicular DC, meaning that these parameters are well constrained by the data, while it is wider for the slopes of the parallel DC, meaning that these parameters are not well constrained by the data.
This is expected, as parallel diffusion dominates the GCR transport in the inner heliosphere, while in the outer heliosphere, where the majority of the modulation takes place, perpendicular diffusion dominates.
These preliminary results are in agreement with what found in \cite{bib:corti19:numericalmodeling} using an ordinary least-square minimization procedure on the same AMS-02 data and numerical model.

\section{Results}

\begin{figure}[!h]
   \centering
   \includegraphics[width=0.96\textwidth]{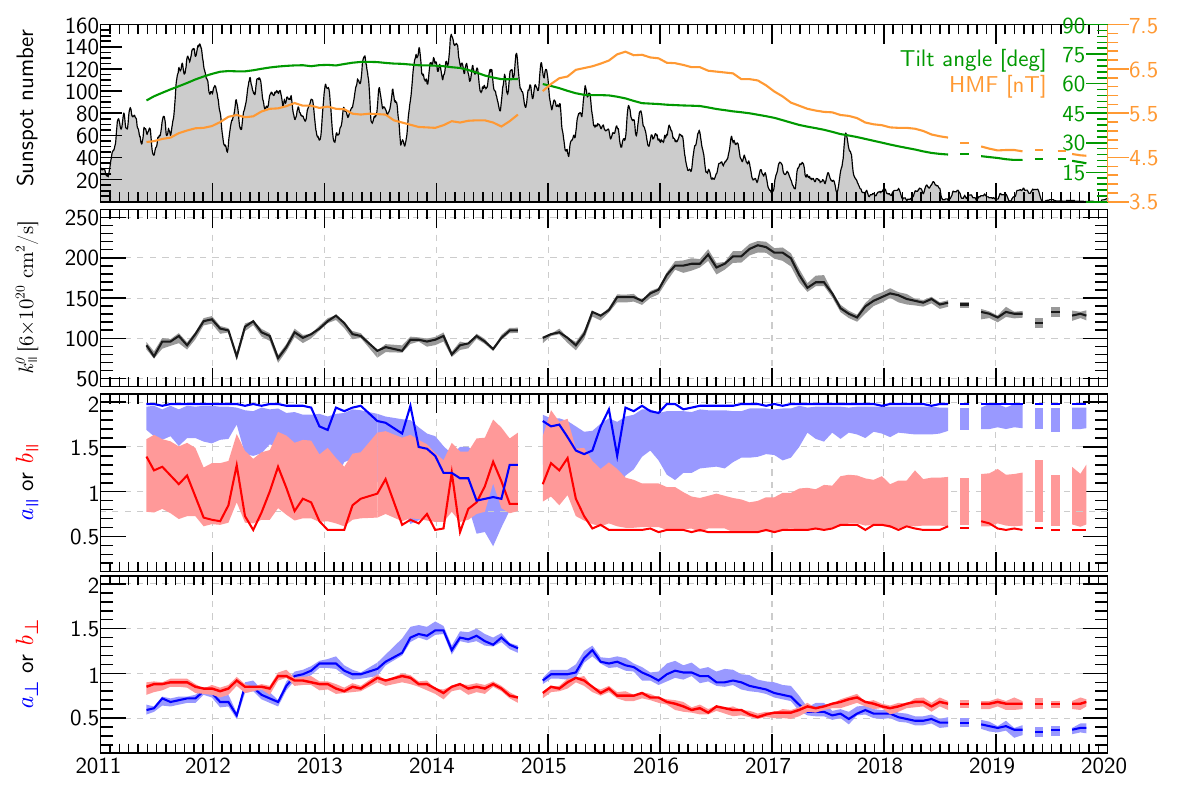}
   \caption{
      Time dependence of the parameters inferred from the HMC.
      For reference, in the first panel the sunspot number and the 1-year backward averaged tilt angle and HMF intensity at 1 AU are shown.
   }
   \label{fig:HMC-pars}
\end{figure}

Figure \ref{fig:HMC-pars} shows the time dependence of the parameters inferred from the HMC on the GCR protons measured by AMS-02 between May 2011 and October 2019, together with their 68\% credible intervals.
The normalization of the DC mostly fluctuates around $600\cdot10^{20}$ cm$^{2}$/s until 2015, when it starts to steadily increase up to a maximum in 2017. After, it slowly decreases and become constant in 2019--2020, around a level roughly 40\% higher than in 2011--2014.
The slopes of the parallel DC are not very much constrained over all the time range, but they seem to be constant in time, except possibly during 2014 and 2015.
The slopes of the perpendicular DC are instead very well constrained and have a different time dependence: the slope above 5 GV increases from 2011 to 2014, then decreases until 2019--2020 to a lower value than in 2011; the slope below 5 GV is mostly constant until mid 2013, decreases with various fluctuations until 2017, and then rise again till 2018 to a lower value than in 2011--2013.

According to quasi-linear theory~\citep{bib:jokipii66:qlt}, the rigidity slope of the parallel DC is related to the slope of the HMF power spectrum: $P(k) \propto k^{-\alpha} \longrightarrow \lambda_{\parallel} \propto R^{2-\alpha}$, where $k$ is the HMF spectrum wave number and $\lambda$ the diffusion mean free path.
Similarly, according to non-linear guiding center theory~\citep{bib:shalchi04:nonlinear-parallel-perpendicular}, there is a relation between the rigidity slope of the perpendicular DC and the slope of the HMF power spectrum: $\lambda_{\perp} \propto \lambda_{\parallel}^{\beta} \longrightarrow \lambda_{\perp} \propto R^{\beta(2-\alpha)}$.
Typically, the HMF spectrum is a broken power-law in $k$: $k^{-\alpha_{H}}$ at low wave numbers corresponding to high rigidities; and $k^{-\alpha_{L}}$ at high wave numbers corresponding to low rigidities, with $\alpha_{L} > \alpha_{H}$, such that the power spectrum falls at larger spatial scales.
This would imply $a_{\perp} < b_{\perp}$.
However, as shown in Figure \ref{fig:HMC-pars}, we observe $a_{\perp} > b_{\perp}$ between 2013 and 2017, \textit{i.e.}, during the solar maximum and decreasing phase of SC24.
Taken at face value, this result would suggest that the HMF power spectrum is \emph{increasing} at larger spatial scales, which is very unlikely.
This indicates either a limitation of the steady-state approach during solar maximum, or a non trivial relation between the HMF power spectrum and the rigidity dependence of the mean free path different from what turbulence theory predicts, or a combination of both.

Song \textit{et al.} reproduced PAMELA and AMS-02 GCR protons with a 3D time-dependent SDE model~\citep{bib:song21:numerical-study-solar}, finding $a_{\perp} > b_{\perp}$ before 2011 and after 2016, \textit{i.e.}, during the SC23/24 solar minimum and decreasing phase of SC24.
A more careful comparison of the model ingredients is needed to understand this discrepancy.

\bibliographystyle{JHEP}
\bibliography{abbrs.bib,refs.bib}

\end{document}